%% file: main.tex
\documentclass[conference]{IEEEtran}
\IEEEoverridecommandlockouts

\usepackage{amsmath,amssymb,amsfonts}
\usepackage{algorithm}
\usepackage{algorithmic}
\usepackage{graphicx}
\usepackage{textcomp}
\usepackage{xcolor}
\usepackage{comment}
\usepackage[hidelinks]{hyperref}
\usepackage[capitalise]{cleveref}
\usepackage{siunitx}
\usepackage[backend=biber,block=ragged,style=ieee]{biblatex}
\def\BibTeX{{\rm B\kern-.05em{\sc i\kern-.025em b}\kern-.08em T\kern-.1667em\lower.7ex\hbox{E}\kern-.125emX}}
\AtEveryBibitem{
 	\clearfield{url}
 	\clearfield{doi}
\ifentrytype{inproceedings}{
 	\clearlist{address}
 	\clearlist{publisher}
 	\clearname{editor}
 	\clearlist{organization}
 	\clearfield{pages}
 	\clearlist{location}
	\clearfield{volume}
	\clearfield{series}
	\clearfield{eprint}
	\clearfield{eprinttype}
 }{}
 }

\begin{document}

\title{Collaborative Safe Bayesian Optimization\\
\thanks{This work has been performed with support of Sweden's Innovation Agency.}
}

\author{
Alina Castell Blasco, Maxime Bouton 

\\ \textit{Ericsson Research, Stockholm}
}

\maketitle

\input{abstract}
\input{introduction}
\input{relatedwork}
\input{background}
\input{problemstatement}
\input{algorithm}
\input{experiments}
\input{conclusions}

\printbibliography[heading=bibintoc]
\input{appendix}

\end{document}

%% file: abstract.tex
\begin{abstract}
Mobile networks require safe optimization to adapt to changing conditions in traffic demand and signal transmission quality, in addition to improving service performance metrics. With the increasing complexity of emerging mobile networks, traditional parameter tuning methods become too conservative or complex to evaluate. For the first time, we apply safe Bayesian optimization to mobile networks. Moreover, we develop a new safe collaborative optimization algorithm called \textsc{CoSBO}, leveraging information from multiple optimization tasks in the network and considering multiple safety constraints. The resulting algorithm is capable of safely tuning the network parameter online with very few iterations. We demonstrate that the proposed method improves sample efficiency in the early stages of the optimization process by comparing it against the \textsc{SafeOpt-MC} algorithm in a mobile network scenario.
\end{abstract}


%% file: introduction.tex
\section{Introduction}
Modern mobile networks consist of complex systems where key parameters, such as antenna tilt or beamwidth, directly influence the coverage quality experienced by users. Optimizing these parameters is essential for maintaining efficient network performance and adapting to changing environmental conditions. However, evaluating the impact of new configurations is complex. It requires simulations or testing with real-world systems which might take days and can be challenging due to safety constraints such as maintaining minimum coverage levels or minimum signal quality.

This paper addresses the problem of safely and efficiently optimizing performance-related parameters in mobile network systems, since the evaluation of each new parameter might take up to a day to complete. The goal is to develop a method that maximizes performance subject to safety constraints through online optimization, that is, by safely exploring parameter values directly in the network. In addition, our objective is to use data from collaborative agents to improve the sample efficiency of current methods.

With the increasing complexity of modern mobile networks, such as 5G and future 6G networks, traditional methods like rule-based strategies have become too conservative. Promising approaches like reinforcement learning (RL) require extensive offline training data or high-quality simulators, and safe evaluation in the network is challenging. Bayesian Optimization (BO), and in particular safe Bayesian Optimization (safe BO), addresses some of these limitations. Unlike RL-based methods that require training, safe BO is an online optimization approach. It uses probabilistic models, typically Gaussian processes (GPs), to explore the parameter space and iteratively update the knowledge about performance while ensuring that each new evaluation is safe. Safe BO uses fewer evaluations than RL methods. However, it requires an initial safe set which might be small or poorly located, hence increasing the number of evaluations needed to reach an optimal configuration, or causing the algorithm to be stuck local optima.
We propose a new algorithm called Collaborative Safe Bayesian Optimization (CoSBO) that builds on the standard safe BO algorithm for multiple constraints \textsc{SafeOpt-MC} \cite{berkenkamp2023bayesian}, introducing a collaborative initialization strategy. The proposed method differs from previous methods in several ways. Unlike rule-based systems, it adapts dynamically to the data. In contrast to RL-based methods, it optimizes online with no pre-training required. Although BO already improves efficiency by balancing exploration and exploitation, we further improve sample efficiency by integrating external data into the optimization process and show that it helps discovering new optimal regions. 

The first contribution of this paper is CoSBO, the new algorithm that takes advantage of the similarity between agents, such as antennas in similar geographical or network traffic conditions, to enrich the initial estimation model of the network performance. By identifying the most correlated collaborator among a set of network optimization problems and transferring a subset of its data, the optimization process starts with a more informed model, without requiring additional evaluations at the initial stage. The quality and availability of the collaborator data are key factors to consider, since the benefit may be reduced if the selected collaborator is poorly correlated. The second contribution consists of applying a safe Bayesian optimization method to the telecommunications domain for the first time. This is done through a careful modeling of safety and performance functions for mobile networks and is evaluated through series of simulation-based experiments. The method is shown to reach high-performing configurations with very few parameter evaluations in the network while maintaining formal safety guarantees. In general, this work contributes to the development of adaptive, data-efficient, and safety-constrained optimization techniques for modern mobile networks. 

The outline of this paper continues with related work and background, followed by a description of the problem and the CoSBO algorithm, and finishes with details on the experiments performed, results obtained, and concluding remarks. The code for the algorithm is available at: \url{https://github.com/EricssonResearch/collaborative-safe-bo}, including hyperparameters and simulated datasets used in our experiments.

%% file: relatedwork.tex
\section{Related Work}\label{sec:related-work}
Recent studies provide RL-based approaches that optimize parameters in mobile networks in a multi-agent setting, such as antenna tilt optimization \cite{mendo2023multi}. These methods require an offline training phase, using a large number of simulated data, since it is too risky and expensive to retrieve samples from real systems, and lack safety guarantees. In contrast, our work focuses on online optimization, aims to reduce the number of evaluations to limit risks and computational cost of training, and considers safety guarantees.

Instead of RL-based methods, we rely on \citeauthor{berkenkamp2023bayesian}'s algorithm, \textsc{SafeOpt-MC} \cite{berkenkamp2023bayesian}, in the core structure of our method. This algorithm balances exploration of the safe set of configurations and exploitation of the optimal one, similar to \citeauthor{sui2015safe}'s algorithm, \textsc{SafeOpt} \cite{sui2015safe}. \citeauthor{maggi2021bayesian} \cite{maggi2021bayesian} applied Bayesian optimization to radio resource management in cellular networks. However, their approach does not account for safety constraints during optimization nor collaborative data. In this paper, we use \citeauthor{berkenkamp2023bayesian}'s approach that formalizes performance and safety as separate functions to optimize and adapt it to mobile network scenarios.

Compared to \textsc{CoSBO}, none of the safe BO algorithms provides a collaborative framework. Existing multi-agent optimization works create simulated functions \cite{lubsen2024towards}, or optimize multiple agents in parallel \cite{chen2024multi}. By contrast, we work with real-world performance functions, and we aim to safely optimize a single agent, respectively. Other works \cite{zhu2020multi} \cite{vinod2022safe} define a global reward function and a global safety constraint, which is useful for planning instead of optimization. Our method uses similarities between different parameters and network areas to both speed up the optimization of the safe BO method and broaden the exploration while maintaining safety guarantees.

%% file: background.tex
\section{Background}\label{sec:background}

In this section, we present Gaussian processes (GPs) and the standard safe Bayesian optimization method, which are the mathematical foundations of our Collaborative Safe Bayesian optimization algorithm in \cref{sec:algorithm}. 

\subsection{Gaussian Processes (GPs)} \label{GPs}
Optimization problems involve finding the best parameter configuration to achieve the maximum possible performance. Generally, this performance or objective function is unknown, so we use models that can approximate the behavior of the system based on previous observations. These approximations are called probabilistic surrogate models, they construct estimations \begin{math}\hat f\end{math} of the true objective function \begin{math}{f}\end{math} based on observations or evaluations of this function and quantify confidence in these predictions \cite{kochenderferwheeler2019}. 

We use Gaussian processes (GPs) as our probabilistic surrogate model since it is suitable for complex black-box functions and can explicitly account for noisy performance measurements, with the predictive variance reflecting this uncertainty \cite{berkenkamp2023bayesian}.
GPs represent a probability distribution over functions, i.e., the estimated function values are modeled as samples of a Gaussian distribution \cite{rasmussenwilliams2006gaussian}. That is, the predicted value at any point in the function comes with an uncertainty estimate. This predicted point represents the point of the true objective function, and this confidence is used to decide which point to evaluate next. GPs construct a GP \textit{prior distribution} when no evaluations have been done on the objective function, this distribution is \begin{math}f(\mathbf{x}) \sim \mathcal{GP}(\mu(\mathbf{x}), k(\mathbf{x}, \mathbf{x}')),\ \mathbf{x}, \mathbf{x}' \in \mathcal{X}, \end{math} where \begin{math}\mu(\mathbf{x})\end{math} is the prior \textit{mean function} and \begin{math}k(\mathbf{x}, \mathbf{x}')\end{math} is the \textit{covariance function} or \textit{kernel} between two points \begin{math}\mathbf{x}, \mathbf{x}'\end{math} of \begin{math}\mathcal{X}\end{math}. In this work, the parameter space \begin{math}\mathcal{X}\end{math} corresponds to domains of network configuration parameters such as the electrical tilt of an antenna. The mean function represents prior knowledge about the function and the kernel controls the smoothness of the functions \cite{kochenderferwheeler2019}. The kernel we use is the squared exponential (RBF), formally described in \cite{kochenderferwheeler2019}, eq. (15.9), due to its particular smoothness properties and common usage in the safe optimization literature. Given that we have a set of points \begin{math}\mathbf{x}\end{math} and their associated observations \begin{math}\mathbf{y}\end{math}, the next step is to predict the value \begin{math}\mathbf{\hat y}\end{math} at point \begin{math}\mathbf{x^*}\end{math} conditioned on the previously observed values. Thus, we construct the GP \textit{posterior distribution}, the distribution of possible unobserved values conditioned on observed values \cite{kochenderferwheeler2019}.

\subsection{Safe Bayesian Optimization} \label{sec:safebayopt}
Safe BO algorithms aim to maximize an unknown objective function while ensuring that every selected parameter satisfies safety constraints with high confidence. The optimization is restricted to a safe set of parameters that is gradually expanded as the algorithm gains more information about the behavior of the system. The algorithm starts with an initial safe set of points \begin{math}S_0 \subseteq \mathcal{X}\end{math} where safety constraints are satisfied. It then iteratively explores the parameter space to expand the region classified as safe while balancing exploitation to find the optimal value. The point of convergence is one reachable from the initial safe region through safe steps. Thus, if the initial set is poorly defined and the exploration hyperparameter is conservative, the algorithm may fail to explore beyond it and will converge to the optimal point within the safe region explored \cite{berkenkamp2023bayesian} \cite{sui2015safe}.

%% file: problemstatement.tex
\section{Problem Statement}\label{sec:problem}
In this section, we present a formal definition of the constrained optimization for mobile networks and the system model.

\subsection{Constrained Bayesian optimization}
Our goal is to maximize an unknown performance function \begin{math}f: \mathcal{X} \rightarrow  \mathbb{R},\end{math} \textit{e.g.} throughput per user or signal quality of an antenna, where \begin{math} \mathcal{X} \subseteq \mathbb{R}^d \end{math} is the domain of input parameters such as electrical tilt. Since \begin{math}f\end{math} is not known \textit{a priori}, we can only learn about it through evaluations at selected parameter points \begin{math} \mathbf{x} \in \mathcal{X} \end{math}, \textit{e.g.} certain tilt values, which yield observations that are noisy because signal propagation is influenced by diverse environmental and network factors, of the form: 
\begin{align} \label{eq:fhat}
\hat f(\mathbf{x} ) = f(\mathbf{x} ) + \omega_n, \quad \omega \sim \mathcal{N}(0,\sigma^2_n),
\end{align} where \begin{math}\omega_n\end{math}  is a Gaussian noise with zero mean and \begin{math}\sigma^2_n\end{math} noise variance. The variance quantifies the uncertainty of the predictions.
To optimize our objective function \begin{math}f\end{math} is to obtain its maximum value. For that, we construct an accurate estimate function \begin{math}\hat f\end{math}, or surrogate model, based on these noisy observations. Specifically, we sequentially evaluate a finite set of parameter points \begin{math}X = [\mathbf{x}_1, \mathbf{x}_2, ..., \mathbf{x}_n]\end{math} and obtain corresponding noisy observations \begin{math}\hat f(\mathbf{x}_i)\end{math} for each \begin{math}\mathbf{x}_i \in X\end{math}. The surrogate model leverages these observations to approximate the true underlying function \begin{math}f\end{math}, guiding the search for its maximum. The evaluated parameter points must satisfy a set of safety constraints (e.g., fraction of users above a signal quality level) defined by \begin{math}q\end{math} unknown constraint functions of the form \begin{math} g_i(\mathbf{x} ) \geq 0,\ g_i: \mathcal{X} \rightarrow \mathbb{R}\end{math}, where \begin{math} i \in\mathcal{I}_g = \{1, \dots , q\} \end{math} are the safety function indices that belong to the set of function indices \begin{math} \mathcal{I}_g \subset \mathcal{I}\end{math}. Similarly as the objective function, the safety constraints are estimated from noisy observations since they are unknown \textit{a priori}. Neither the performance function nor the safety constraints are known. However, we assume the existence of an initial safe set of parameters \begin{math}S_0 \subseteq \mathcal{X}\end{math} known to satisfy the safety conditions. This starting point is derived from domain knowledge. In telecommunications applications, prior experience and system knowledge allow experts to identify parameter configurations that are guaranteed to be safe, for example, based on current deployment practices. With our \textsc{CoSBO} algorithm we provide an initial safe set intelligently informed from experience of different antenna parameter tuning problems.

The optimization process then proceeds iteratively: at each time step \begin{math}t\end{math}, the algorithm selects a new parameter point \begin{math}\mathbf{x}_t \in \mathcal{X}\end{math} to evaluate, ensuring that \begin{math}\mathbf{x}_t\end{math} is safe according to the current model estimates. Only parameters within the feasible region (where all safety constraints are satisfied) are evaluated. Formally, the optimization problem is defined as:
\begin{align}\label{eq:optprobideal}
    \max_{\mathbf{x} \in \mathcal{X}} f(\mathbf{x}) \; \text{ subject to } \; g_i(\mathbf{x}) \geq 0, \;i=1, \dots, q.
\end{align}

Nevertheless, as mentioned in \cref{sec:safebayopt}, we can only find the optimal point reachable from the initial safe region and learn the safety constraints up to some accuracy  \begin{math}\epsilon\end{math}. Therefore, we define the reachable safe set \begin{math}R_{\epsilon}(S_0)\end{math} as the set of parameters that can be safely explored starting from the initial set \begin{math}S_0\end{math}, given the estimation accuracy \begin{math}\epsilon\end{math}, and without going through unsafe parameters. Consequently, the true optimal value is defined as:
\begin{align} \label{eq:opt}
    f^*_\epsilon = \max_{\mathbf{x} \in R_\epsilon(S_0)} f(\mathbf{x}) \text{ s. t. }\; g_i(\mathbf{x}) \geq 0, \;i=1, \dots, q ,
\end{align}
where
\begin{align} \label{eq:reachable}R_\epsilon (S) := S \cup \bigcap_{i \in \mathcal{I}_g} \{\mathbf{x} \in \mathcal{X} \; | \; \exists \;\mathbf{x}' \in S : g_i (\mathbf{x}') - \epsilon \geq 0\}\end{align}

\subsection{System Model}
Having introduced the formal problem, we now describe a real mobile network application. This scenario takes place in a region composed of multiple urban areas, where each has its own size, population density, geographical topology, and mobile network infrastructure with base stations and antennas. These factors influence the performance of the antennas and are reflected on their performance functions which may differ or show similarities among areas with comparable characteristics. The optimization problem begins when a new antenna is to be deployed: its performance function is initially unknown and must be optimized to identify the best configuration, while simultaneously satisfying a safety constraint that guarantees minimum coverage for the served cell.

\begin{figure}[t]
    \begin{center}
    \includegraphics[width=\linewidth]{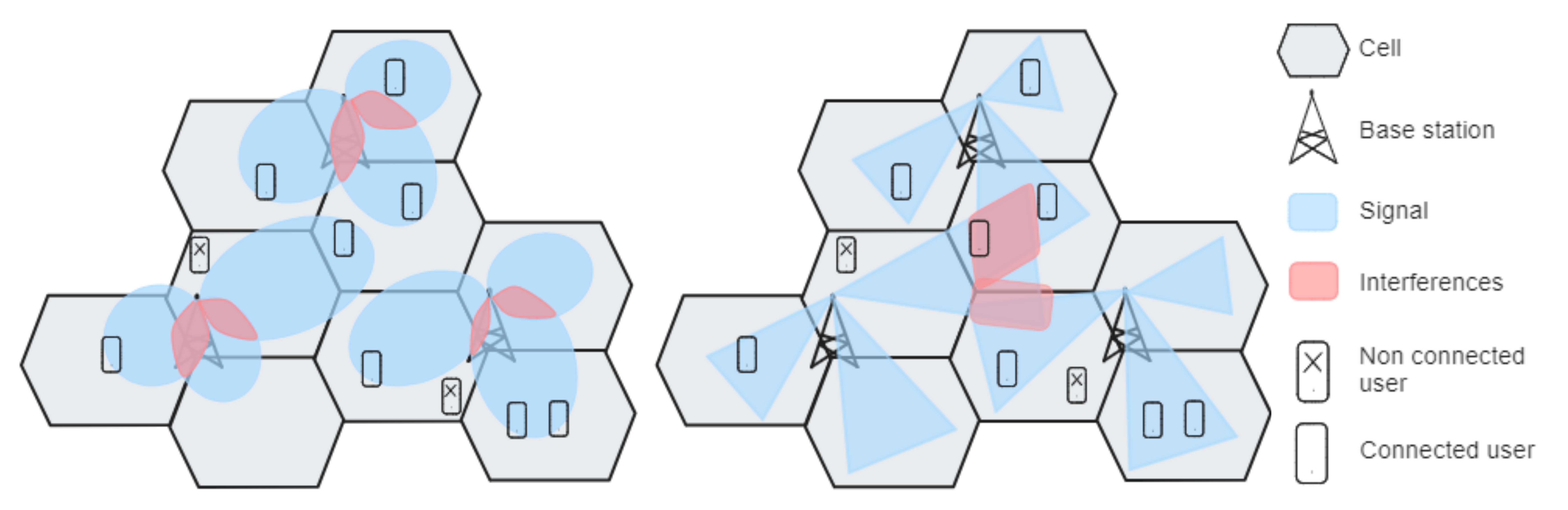}
    \caption{Illustration of two mobile network scenarios where the antenna parameter being adjusted on the left is the horizontal beamwidth, and on the right is the electrical tilt. Each base station includes three cells. Both scenarios show two negative outcomes (non connected users) resulting from beamwidth/tilt configuration and insufficient coordination.}
    \label{fig:bw_tl}        
    \end{center}
\end{figure}

We consider two performance functions (\begin{math}f\end{math}) corresponding to different parameters illustrated in \cref{fig:bw_tl}, beamwidth and tilt, in addition to a coverage safety constraint (\begin{math}g\end{math}). We define the following measurements based on \textit{Reference Signal Received Power} (RSRP) since it is proportional to the antenna gain and the antenna gain is a function of tilt and beamwidth as described in \cite{farooq2019ai}, eq. (3). We compute the \textit{Signal over Interference and Noise Ratio} (SINR) according to \cite{farooq2019ai}, eq. (4), and the throughput according to \cite{farooq2019ai}, eq. (5). The beamwidth performance \begin{math}f\end{math} is quantified by the number of users with good coverage and low interference, the tilt performance \begin{math}f\end{math} by the average user throughput and the safety constraint \begin{math}g\end{math} by the number of users with acceptable RSRP. The respective formulas are as follows:
\begin{align} \label{eq:bw}
\text{TotalUsers} = \sum_{u=1}^N \mathbb{I}\big[ &\, \text{Interference}_u < h_{\text{interference}}  \notag\\
&\wedge \; RSRP_u > h_{\text{RSRP}} \big]
\end{align}
\begin{align}
\text{Throughput} &= \frac{1}{N} \sum_{u=1}^N \text{Throughput}_u  \label{eq:tl}\\
\text{RSRP}_{\text{count}} &= \sum_{u=1}^N \mathbb{I}\big[\text{RSRP}_i > h_{\text{RSRP}} \big], \label{eq:coverage}
\end{align}

where \begin{math}u\end{math} corresponds to users, $N$ is the total number of users, and the thresholds \begin{math}h_{\text{interference}}\; \text{ and } \; h_{\text{RSRP}}\end{math} thresholds correspond to the 50th percentile of interference and 5th percentile of RSRP across users in dBs, respectively, measured in the initial state and used as constants throughout the optimization. 

%% file: algorithm.tex
\section{The \texttt{CoSBO} method}\label{sec:algorithm}
In this section, we introduce a novel collaborative extension of the \textsc{SafeOpt-MC} algorithm \cite{berkenkamp2023bayesian}. Our algorithm leverages multiple agents by identifying the one most similar to the main function and transferring its data to the main agent, thereby enhancing the optimization process. We illustrate an example run in \cref{fig:bw_iterations}, and the full procedure is described in \cref{alg:cosbo}.

\subsection{Collaborative initialization strategy}

We propose a collaborative strategy to enhance the sample efficiency of \textsc{SafeOpt-MC} \cite{berkenkamp2023bayesian} by using information from external agents, achieved by incorporating observations from collaborator agents into the main agent's GP model during initialization. The key is to identify the collaborator whose behavior most closely correlates with the main agent by calculating the similarity between the main agent and the collaborators. However, since at initialization there is no direct data available from the main agent, we propose to use an adjacent optimization domain, as described next.

We define \begin{math}\mathcal{X}_A\end{math} as the \textit{main} set of input parameters (e.g., electrical tilt) over which we optimize the main agent (e.g., an antenna). Then, we assume that there exists another domain, denoted \begin{math}\mathcal{X}_B\end{math}, over a separate set of input parameters (e.g., horizontal beamwidth). Additionally, we assume that all agents, both collaborators and the main agents, have estimated objective functions on this domain and we have access to them. This assumption could easily be satisfied by starting a safe optimization on one parameter that we know is safer, as we demonstrate using safe BO. Alternatively, one could use the history of network configuration changes performed on the area being optimized. 

Afterwards, we compute the Pearson correlation \cite{navarro2025pearson} between the data points of the collaborators and the main agent on domain \begin{math}\mathcal{X}_B\end{math} (e.g., horizontal beamwidth), formally:
\begin{align} \label{eq:pearson}
    \rho(\mathbf{x}_{\mathcal{X}_B}, \mathbf{x'}_{\mathcal{X}_B}) = \frac{\mathbb{E}[(\mathbf{x} - \mu_\mathbf{x})(\mathbf{x'}- \mu_\mathbf{x'})]}{\sigma_{\mathbf{x}} \sigma_{\mathbf{x'}}},
\end{align}
where \begin{math}\mathbf{x}, \mathbf{x'} \in \mathcal{X}_B\end{math} represent the main agent and collaborator agents, and \begin{math} (\mu_\mathbf{x},\sigma_{\mathbf{x}})\end{math} and \begin{math} (\mu_\mathbf{x'},\sigma_{\mathbf{x'}})\end{math} correspond to their statistical estimates, respectively. Next, the collaborator with highest correlation is identified and, on the main input domain \begin{math}\mathcal{X}_A\end{math}, we select the \begin{math}k\end{math} most correlated samples, by default \begin{math}k=10\end{math}, from its GP posterior distribution. These samples are then used to initialize the GP model of the main agent for optimization. Importantly, while the correlation is computed using \begin{math}\mathcal{X}_B\end{math}, the transferred data originates from \begin{math}\mathcal{X}_A\end{math} under the assumption that pairwise similarity across collaborators in \begin{math}\mathcal{X}_B\end{math} reflects that in \begin{math}\mathcal{X}_A\end{math}. In other words, two network areas that show strong correlation when being optimized for beamwidth will most likely be correlated for other parameters as well. This is proven to hold for our simulation experiments in \cref{sec:experiments_results}.

\subsection{The \textsc{CoSBO} algorithm}\label{sec:algorithm_section_B}
\Cref{alg:cosbo} starts by fitting the GP models from a known safe initial parameter value \begin{math}\mathbf{x}_0\end{math} (e.g., a tilt value) and its measures \begin{math} \hat f(\mathbf{x}_0), \hat g_i(\mathbf{x}_0)\end{math} (e.g., throughput per user and number of users with acceptable RSRP). Then, selects the most correlated collaborator using the Pearson correlation \begin{math}\rho(\hat f_{\mathcal{X}_B}, \hat{f}_{C_{\mathcal{X}_B}})\end{math} between the main agent's GP posterior and the collaborator agents from the set \begin{math}C_{\mathcal{X}_B}\end{math}. After that, it selects the collaborator data from the main set \begin{math}C_{\mathcal{X}_A}\end{math} and identifies the \textit{k} most informative data points (the most correlated ones), which are stored in the set \begin{math}\mathcal{D}_k\end{math}. Then, updates the \begin{math}\mathcal{GP}\end{math} model of the main agent's performance and safety functions with high-quality prior information. The optimization process then proceeds with the \textsc{SafeOpt-MC} \cite{berkenkamp2023bayesian} algorithm. 

\begin{figure*}[t]
    \begin{center}
    \includegraphics[width=\textwidth]{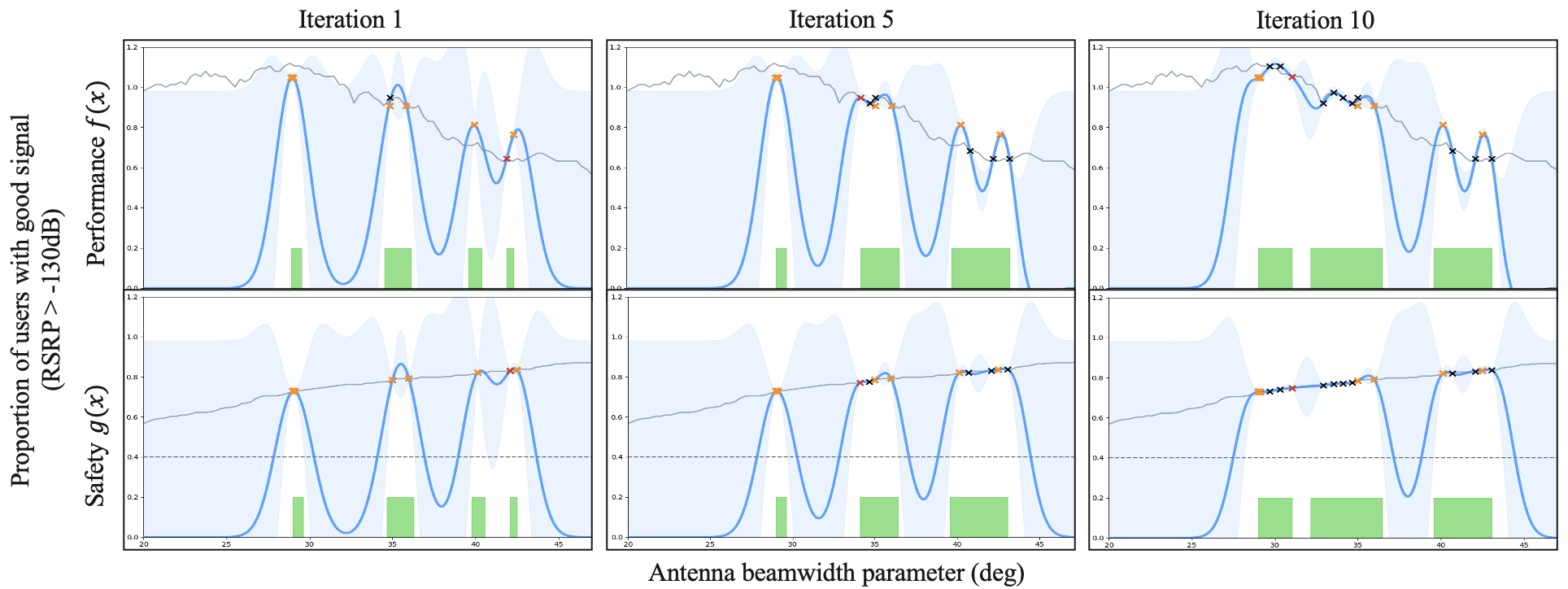}
    \caption{Optimization process of the horizontal beamwidth of an antenna using the CoSBO algorithm to maximize the performance function (gray curve on the top) while satisfying the safety constraint (gray curve on the bottom). The visualization shows the progress of the estimated functions (light blue curves) at the first, fifth, and tenth iterations. The algorithm starts with collaborator data (orange crosses) that is added as observed data (black crosses). Then, based on the GP posterior and its confidence intervals (light blue areas), it iteratively selects and evaluates new data (red crosses) that are above the safety threshold (gray dashed line) and within the current safe set (green set). The input parameter region is truncated on the plot.}
    \label{fig:bw_iterations}        
    \end{center}
\end{figure*}

The visualization in \cref{fig:bw_iterations} represents the optimization process of the horizontal beamwidth parameter of an antenna that aims to maximize the number of users with good coverage and low interference (performance function \begin{math}f\end{math}) while ensuring a minimum number of users with acceptable RSRP (safety constraint \begin{math}g\end{math}). The algorithm starts with the estimation of both the performance and safety function based on the transferred data from a collaborator (orange crosses). With the GP (light blue curve) and the confidence intervals (light blue areas), the algorithm is able to determine current observations and nearby parameter points as safe (green set). At each iteration, an evaluation (red cross) of the real objective function (gray curve) is done and added to the GP posterior. The selected point to evaluate belongs to the safe set and is chosen because it either expands this safe set or maximizes the performance. The safety threshold (gray dashed line) indicates that the points below it are unsafe. In this example, the addition of collaborative data has created three separate safe regions, the algorithm takes advantage of this to explore all of them; by increasing the safe set of points, we improve the sample efficiency of the process.  

\begin{algorithm}[t]
\caption{CoSBO}\label{alg:cosbo}
\begin{algorithmic}[1]

 \REQUIRE \begin{tabular}[t]{@{}l}
 Initial safe parameter \begin{math}\mathbf{x}_0 \in S_0 \subseteq \mathcal{X}_A\end{math},\\
 \begin{math}\mathcal{GP}\end{math} priors \begin{math}((\mu_0, \sigma_0, \mathbf{k}), (\mu_i, \sigma_i, \mathbf{k}_i)), \; \forall i \in \mathcal{I}_g\end{math}, \\
 \begin{math} \mathcal{GP} \end{math} posterior \begin{math}\hat f_{\mathcal{X}_B} \end{math}, \\
 collaborator sets \begin{math}C_{\mathcal{X}_A}, C_{\mathcal{X}_B} \end{math}
\end{tabular}

 \STATE Initialize \begin{math} \mathcal{GP} \gets \mathbf{x}_0, \hat{f}(\mathbf{x}_0), \hat{g}_i(\mathbf{x}_0),\; \forall i \in I_g
 \end{math}
 \STATE Identify most correlated collaborator: \begin{math}\mathbf{z}_c = \rho(\hat f_{\mathcal{X}_B}, \hat{f}_{C_{\mathcal{X}_B}})\end{math}
 \STATE Collaborator data: \begin{math}(\hat f(\mathbf{x}), \hat g_i(\mathbf{x}))\in C_{\mathcal{X}_A}, \forall i \in \mathcal{I}_g \end{math}
 \STATE \begin{math}\mathcal{D}_k \gets \{ \text{top } k \text{ points from collaborator data} \} \end{math}
 
 \FOR{d = 1, \dots, k}
 \STATE Let \begin{math}x_d \in \mathcal{D}_k \end{math}
 \STATE Update \begin{math} \mathcal{GP} \gets \mathbf{x}_d, \mathbf{z}_c, (\hat{f}(\mathbf{x}_d), \hat{g}_i(\mathbf{x}_d))\in C_{\mathcal{X}_A},\; \forall i \in I_g
 \end{math}
 \ENDFOR

 \FOR{t = 1, \dots }
 \STATE Run \textsc{SafeOpt-MC}
 \ENDFOR
\end{algorithmic}
\end{algorithm}

The data points obtained from the optimization process are evaluations of the real objective function, therefore, the GP models should have more confidence on those points than on the transferred ones from the collaborator. This is formally defined by extending the performance function to incorporate a \textit{context} variable \begin{math}f: \mathcal{X} \times \mathcal{Z} \rightarrow \mathbb{R}\end{math}. Therefore, for each input \begin{math}\mathbf{x} \in \mathcal{X}\end{math} the GP model is augmented with an associated context value \begin{math}\mathbf{z} \in \mathcal{Z}\end{math} and measurements are of the form \begin{math}\hat f(\mathbf{x}, \mathbf{z})\end{math}~\cite{krause11}. On the one hand, the context value associated to the collaborator data is the correlation coefficient between the main agent and the collaborator, formally, \begin{math}\rho(\mathbf{x}_{\mathcal{X}_B}, \mathbf{x'}_{\mathcal{X}_B}) =\mathbf{z}_c \in \mathcal{Z}
\end{math}. On the other hand, the value associated to the measurements from the main agent is \begin{math}\mathbf{z}_m = 1\end{math}, the maximum Pearson correlation coefficient which is 1 \cite{navarro2025pearson}. Hereby, the transferred data is of the form \begin{math}(\mathbf{x}, \mathbf{z}_c)\end{math} and is quantified with a higher uncertainty than the data evaluated from the real objective function \begin{math}(\mathbf{x}, \mathbf{z}_m)\end{math}.

Quantifying the context is essential to prevent low-quality collaborators from degrading performance, since all transferred samples are assigned lower confidence than evaluations obtained directly from the main agent. Consequently, if a collaborator has misleading correlations, its influence decreases rapidly as real evaluations are accumulated. In the worst-case scenario, where none of the transferred data points contribute useful information for identifying the objective function, the effective optimization process begins once the main agent’s own evaluations are incorporated. A safeguard mechanism against misleading correlations is proposed in \cref{sec:conclusions}.

\subsection{The \textsc{SafeOpt-MC} algorithm}
This algorithm guarantees safety, and balances exploration and exploitation; that is, it iteratively selects points that can either expand the safe set of parameters (e.g., explore a region of the network with possibly poorer coverage) or maximize the objective function (e.g., maximize received signal quality). To ensure safety, the algorithm defines confidence bounds around the \begin{math}\mathcal{GP} \end{math} posterior of both \begin{math}f \text{ and } g_i\end{math} with its predictive posterior mean \begin{math}\mu_t(\mathbf{x}) \end{math} and variance  \begin{math}\sigma_t^2(\mathbf{x}) \end{math}, which contain the true functions \begin{math}f \text{ and } g_i \end{math} with high probability \cite{berkenkamp2023bayesian}:
\begin{align}\label{eq:CI}
Q_t(\mathbf{x}) := [\mu_{t-1}(\mathbf{x}) \pm \beta_t^{1/2} \sigma_{t-1}(\mathbf{x})],
\end{align}
where the hyperparameter \begin{math}\beta\end{math} controls the desired confidence level. These confidence intervals are used to compute safe regions and guide the selection of the next query point to evaluate. A point is considered safe if its lower confidence bound \begin{math}l(\mathbf{x})\end{math} remains above the safety threshold \textit{h}. The set of safe points are defined as the safe region:
\begin{align}\label{eq:safeset}
S_t = \bigcap_{i \in I_g} \bigcup_{\mathbf{x} \in S_{t-1}} \{\mathbf{x'}\in\mathcal{X} \; | \; l^i_t(\mathbf{x}) \geq h\},
\end{align}
 which contains all points in the set \begin{math} S_{t-1} \end{math} \cite{berkenkamp2023bayesian}. 
 To trade off between exploration and exploitation, the algorithm identifies two subsets of the safe set at every iteration \begin{math}t \end{math}: the set of potential maximizers (M), that could improve the estimate of the maximum, and potential expanders (E), that could expand the safe region. The maximizers are the safe points whose upper confidence bound (\begin{math}u(\mathbf{x})\end{math}) is higher than the largest lower confidence bound (\begin{math}l(\mathbf{x})\end{math}) of the performance function \begin{math}f \end{math}. The expanders are the safe points that would produce a posterior distribution with a larger safe set if they were added to the \begin{math}\mathcal{GP} \end{math} of the safety functions \begin{math}g_i \end{math}. These points are the most uncertain, providing the most information and naturally lying near the boundary of the safe region \cite{kochenderferwheeler2019}. These sets are originally defined by \citeauthor{sui2015safe} \cite{sui2015safe}.

Every time a new point is added to the GP model, the confidence bounds \textit{Q}, the safe set \textit{S}, the maximizers set \begin{math}M \end{math} and expanders set \begin{math}E\end{math} are updated. This new point \begin{math}\mathbf{x}_t \end{math} is the one with highest predictive uncertainty, thereby balancing safe exploration and performance improvement:
\begin{align} \label{eq:nextpoint}
\mathbf{x}_t = \underset{\mathbf{x} \in E_t \bigcup M_t}{\arg\max} \; \max_{i \in I} w_n(\mathbf{x},i),
\end{align}
\begin{align} \label{eq:w}
w_n(\mathbf{x},i) = u_t(\mathbf{x} ) - l_n(\mathbf{x} )
\end{align}
This iterative process of expanding the safe set of points, exploiting promising regions, and exploring most uncertain ones, continues until no new safe points can be identified or no potential improvement is possible. 

%% file: experiments.tex
\section{Experiments}\label{sec:experiments}
To assess the performance of \textsc{CoSBO} we conduct experiments on a set of simulated mobile network scenarios and compare our algorithm against \textsc{SafeOpt-MC} \cite{berkenkamp2023bayesian}. In particular, we aim to answer the following questions: Does \textsc{CoSBO} improve sample-efficiency? Is using data from a poorly correlated collaborator still more sample efficient than no collaboration at all? 

\begin{figure}[t]
    \begin{center}
    \includegraphics[width=\linewidth]{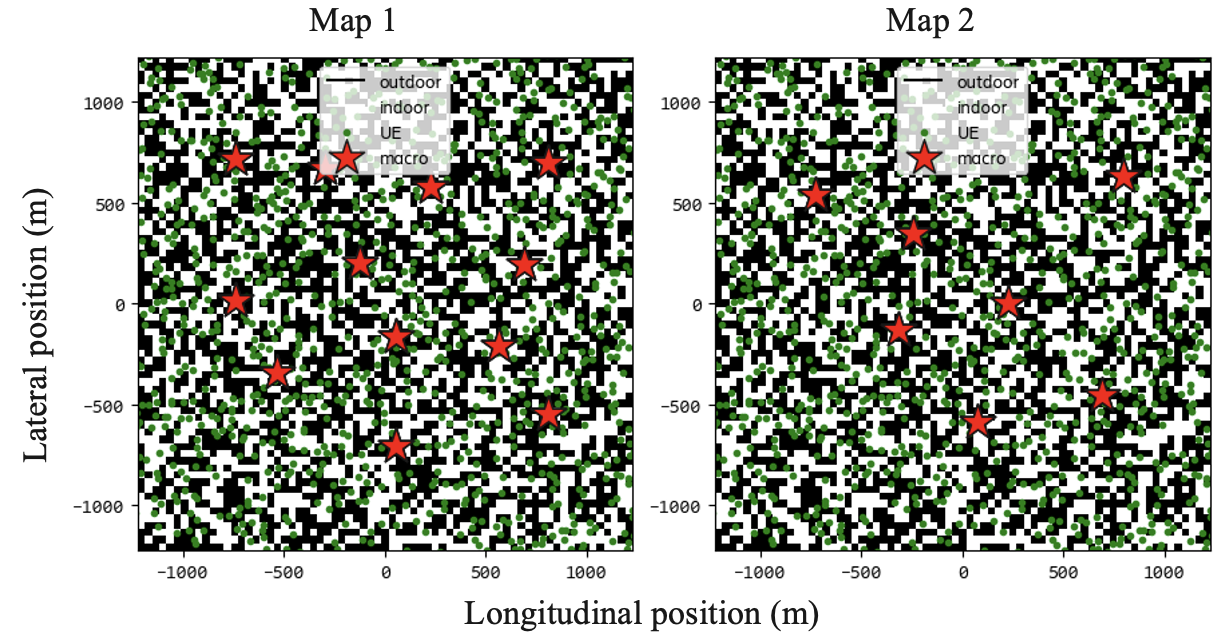}
    \caption{Two of the five simulator-generated topological maps for traffic volumes \SI{15e7}{\bit\per\second}. The mobile network is comprised by base stations (macro, red stars), the users (green dots), the obstacles or indoor areas (white squares), and the outdoor areas (black squares). The indoor-outdoor probability corresponds to a \qty{0.5} quantity.}
    \label{fig:simulatormap}        
    \end{center}
\end{figure}

\subsection{Simulation Environment}
The experiments were carried out using a proprietary system level simulator that simulates realistic behavior of antennas and base stations modeling the signal propagation \cite{asplund2018set}, and using 3GPP models \cite{8005} for NR urban macro scenarios. We constructed five artificial maps to emulate the topology of a city based on a default square layout and placed seven to twelve base stations according to a Poisson point process, providing spatial randomness in their distribution, with a minimum intersite distance. An example of these maps can be seen in \cref{fig:simulatormap}.
In addition to the topological variability of the different maps, we incorporated traffic load variations which represent the amount of data moving across a network. 
We defined three load levels: low, medium and high, correspond to a percentage of utilization of \qty{5}-\qty{10}{\percent}, \qty{30}{\percent} and \qty{60}-\qty{80}{\percent}, and total traffic volumes of \SI{5e7}{\bit\per\second}, \SI{15e7}{\bit\per\second} and \SI{25e7}{\bit\per\second}, respectively. The traffic volume is spread uniformly across uniformly distributed users in the map. All these factors influence the performance and safety functions producing unique network optimization problems for every scenario. With 5 maps and 3 load levels we have a total of 15 different network scenarios. The simulated data is available in our code repository.

The optimization task was conducted over two parameters of the antennas: the horizontal beamwidth and the electrical tilt. These parameters were modeled using a urban macro propagation model with frequency of 2GHz using the standard reference of 3GPP TR 38.921 \cite{8005}. They influence the radiation pattern of the antenna and directly impact user experience metrics. 
The beamwidth determines the angular spread of the antenna's main lobe in the horizontal plane, narrower beamwidths result in poorer signal coverage at the boundary of the cell sector (beam distortion is more likely to occur) while wider ones results in good signal coverage. 
The tilt controls the vertical orientation of the antenna, narrower tilts result in more focused coverage areas, which can increase signal strength but may cause some users to lose coverage; while wider tilts ensure no user loses coverages but also introduce interference if not properly aligned with the rest of antennas.

\subsection{Hyperparameter configuration}
The hyperparameter configuration we used to setup the topological map is described in \cref{tab:maps}. In addition, we used a unique GP regression model to estimate each of the performance and safety functions that also has tunable hyperparameters including kernel function parameters, noise variance, and constants specific for the algorithms \textsc{CoSBO} and \textsc{SafeOpt-MC} \cite{berkenkamp2023bayesian}, all of which are described in \cref{tab:gpconfig}. We tune the values of the variance and lengthscale, which controls the smoothness of the function, how quickly is it expected to change in terms of y values. The chosen value for the lengthscale is \begin{math} \ell = 1\end{math}, as it provides a good balance of smoothness and is a standard choice in the literature. The noise variance is assigned different values for the performance and safety functions to rely more confidently on safety estimates while accounting for potential variability in performance measurements. In the \textsc{SafeOpt-MC} algorithm we use the exploration constant \begin{math}\beta \end{math} which balances exploration and exploitation while preserving safety; we choose the value \begin{math}\beta = 2\end{math} since higher values focus on exploring the parameter space instead of exploiting it, while lower values focus on maximization and get stuck in local optima. In the \textsc{CoSBO} algorithm we define the safety threshold \textit{h}, which determines the safe regions that can be explored. Since the safety constraint corresponds to the number of users with a signal quality level (RSRP) higher than \SI{-130}{\decibel} (\cref{eq:coverage}), the threshold \textit{h} is set to the 50th percentile of users that must have coverage.

\begin{table}[t]
\caption{Topological map parameters.}
    \begin{center}
    \label{tab:maps}
    \begin{tabular}{|c|c|}
    \hline
    \textbf{Square Map} & \\ 
    Edge length & \SI{2000}{\metre} \\ 
    Number of bins & 5000 \\ \hline
    \textbf{Poisson Distribution} & \\ 
    \begin{math}\lambda \;\text{m}^{-2}\end{math} & 
    \SI{4e6}{}\\ 
    Min. intersite distance & \SI{400}{\metre} \\ \hline
    \textbf{Deployment} & \\ 
    Number of users & 1000 \\ 
    Sectors per site & 3 \\ \hline
    \end{tabular}
    \end{center}
\end{table}

\begin{table}[t]
\caption{GP models and CoSBO parameters.}
    \begin{center}
    \label{tab:gpconfig}
    \begin{tabular}{|c|c|c|c|c|c|c|}
    \hline
         Kernel \begin{math}\sigma^2\end{math}& \begin{math}\ell\end{math} & \begin{math}f\end{math} noise \begin{math}\sigma^2\end{math} & \begin{math}g\end{math} noise \begin{math}\sigma^2\end{math} &\begin{math}\beta\end{math} & Threshold \begin{math}h\end{math} & k\\ \hline
         0.5& 1 & \SI{1e-4}{} & \SI{1e-5}{}& 2 & 0.4& 10 \\ \hline
    \end{tabular}
    \end{center}
\end{table}

\begin{figure*}[t]
    \begin{center}
    \includegraphics[width=\textwidth]{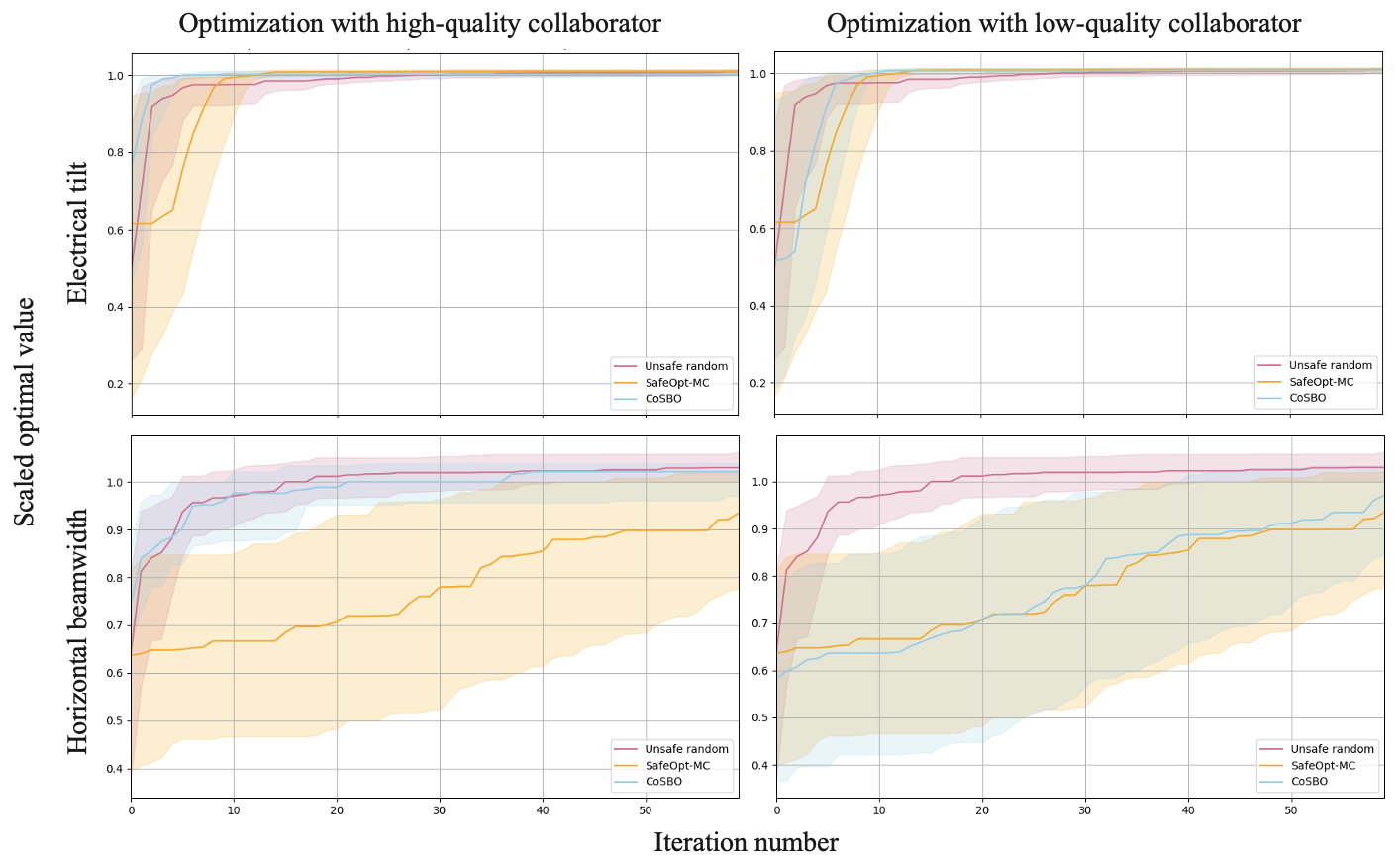}
    \caption{Optimization method comparison for tilt and beamwidth using high-quality collaborators (on the left) and low-quality ones (on the right). Curves correspond to the optimum obtained at each iteration of the process for a budget of 60 iterations. The shaded areas represent the \qty{25}{\percent}-\qty{75}{\percent} inter-quartile range of values obtained over the \qty{15} simulated networks times the \qty{10} different starting points. CoSBO method is displayed in blue, SafeOpt-MC in orange, and an unsafe random exploration method in violet red.}
    \label{fig:optdefaultweak}        
    \end{center}
\end{figure*}

Moreover, we scaled all the experimental scenarios since small differences in the function values can greatly affect the correlation coefficient and values outside the range of values of the GP prior distribution might be less likely to be observed, leading to inaccurate predictions. We applied a Min-Max scaling to all the scenarios using scenario-specific minimum and maximum values to normalize all function outputs to the \begin{math}[0,1]\end{math} interval.
This pre-processing step was possible because the functions were computed from simulations, allowing the true minimum and maximum values to be known beforehand. If those were not known, estimates based on collaborator data or domain-specific knowledge could be used. In practice, historical performance management counters from a live network could be used to perform this normalization step.

\subsection{Fast online optimization in mobile networks}\label{sec:experiments_results}
We measure the sample efficiency of the methods in terms of the highest average value of the objective function found at each iteration in all scenarios of each input parameter, represented in each curve of \cref{fig:optdefaultweak}. The experiments were repeated \num{10} times with different safe initial starting points randomly selected within a safe range. Our algorithm is represented in a blue curve, the baseline algorithm, \textsc{SafeOpt-MC} \cite{berkenkamp2023bayesian}, in an orange curve, and an unsafe random exploration process in violet red. The latter is a method that searches for the optimum by randomly exploring the entire parameter space. Since it does not consider a safety constraint nor a safe region of parameters, it randomly chooses any point from the entire parameter space and rapidly converges towards the true optimal value. Knowing that its convergence is faster than that of the safe optimization algorithms, we use it as a reference for the other two algorithms. 

We separate the results by network parameter being tuned because their performance function values are too different to be plotted jointly. We display the curves with confidence intervals to represent the complete range of values, which are computed over the 15 simulated networks. In the graphic in the upper left of \cref{fig:optdefaultweak}, the \textsc{CoSBO} (blue curve) reaches the asymptotic optimal value in fewer iterations than the \textsc{SafeOpt-MC} (orange curve); this is attributed to the additional data received from the collaborator in the early stages of optimization. In particular, the optimal value of the tilt parameter is reached about seven iterations earlier with the \textsc{CoSBO} algorithm than with \textsc{SafeOpt-MC}, which in a live mobile network deployment can translate into an advantage of seven days in the optimization process since measuring the performance function is often done using daily aggregate of network performance counters \cite{mendo2023multi}. In the bottom left plot of \cref{fig:optdefaultweak}, both algorithms (blue and orange curves) overlap notably with their confidence intervals, meaning that they obtain similar optimal values at each iteration step, still, on average \textsc{CoSBO} obtains higher optimal values during the initial stage of the process. Therefore, the optimization results using the most correlated collaborator (i.e., high-quality) demonstrate that \textsc{CoSBO} is more sample efficient than \textsc{SafeOpt-MC}. 

Furthermore, our method converges with fewer evaluations compared to RL-based approaches in a similar simulation environment. \citeauthor{mendo2023multi} \cite{mendo2023multi} (Figures 6 and 7) rely on the same simulation environment and report that their pre-trained agent requires at least 10 steps to fine-tune to a new network scenario on the tilt optimization problem. Our approach achieves it using fewer steps without the need for pre-training. A comprehensive comparison of BO and RL would significantly widen the scope of this paper. Generally, GP based optimization may scale less than RL in terms of space complexity but requires orders of magnitude fewer samples. We believe the overall trade-off strongly favors BO for the problem at hand. Exploring hybrid RL-BO strategies is an interesting future direction.

We further evaluate the robustness of our collaborative strategy by testing it with lower-quality collaborators, i.e., the least correlated ones. The correlation coefficients of these collaborators is, in average, a \qty{0.39} for the beamwidth parameter functions and a \qty{0.15} for the tilt parameter functions across the \num{15} scenarios. These are significantly low values considering that the maximum Pearson coefficient is \qty{1} and a coefficient of \qty{0} indicates no correlation \cite{navarro2025pearson}. In the plots on the right side of \cref{fig:optdefaultweak} the two compared algorithms perform similarly for both optimized parameters, as indicated by the considerable overlap of their confidence intervals and their convergence to the asymptotic optimal value in similar iterations. These graphics show that when the chosen collaborator is not correlated at all, the algorithm does not gain information. However, the performance of our algorithm does not worsen; in fact, it achieves optimal values comparable to \textsc{SafeOpt-MC} \cite{berkenkamp2023bayesian}. Thus, even under limited collaboration, our algorithm remains as sample-efficient as \textsc{SafeOpt-MC}. This results from the smart use of the context variable computed with the correlation coefficient of the selected collaborator discussed in \cref{sec:algorithm_section_B}. Furthermore, our collaborative strategy relies on the assumption that pairwise similarity across collaborators in domain \begin{math}\mathcal{X}_B\end{math} reflects that in the main domain \begin{math}\mathcal{X}_A\end{math}. The fact that the \textsc{CoSBO} algorithm achieves a higher sample efficiency with high-quality collaborators than with low-quality ones supports this assumption.

%% file: conclusions.tex
\section{Conclusions and Future Work}\label{sec:conclusions}
We presented a collaborative extension of the safe Bayesian optimization framework that enhances the sample-efficiency by leveraging observations collected from other agents deployed in similar, but not identical, environments. The method accelerates the identification of safe and high-performing regions in the parameter space by integrating external data at early stages into the optimization process, i.e., directly into the Gaussian process model. We demonstrated how to model mobile network optimization as a safe Bayesian optimization problem and how to apply \textsc{SafeOpt-MC} \cite{berkenkamp2023bayesian}. Through realistic simulations, we demonstrated that our proposed collaborative strategy improves sample efficiency when using high-quality collaborators. In addition, we prove that our method achieves comparable efficiency to \textsc{SafeOpt-MC} even under low-quality collaborators. \textsc{CoSBO} is particularly effective during the early stages of optimization and recommended in scenarios where safe exploration is essential and the available evaluation budget is limited or complex. Finally, although our experiments rely on simulations, the proposed method could be implemented in a real network within the O-RAN architecture running as an rApp in a centralized non real time Radio Intelligent Controller (RIC)~\cite{oran_arch}. 

Future work includes exploring estimation-based approaches to improve collaborator selection and looking into uncertainty-aware correlation metrics. Extending the telecommunications application to multiple safety constraints which could influence the selection of collaborators is a straightforward extension, as well as expanding the set of parameters to be tuned, which we refrained to do to focus the analysis on the collaborative aspect. Adding a safeguard approach to require a minimum Pearson correlation value for a collaborator to be used. Another avenue of future work could be to combine the collaborative safe exploration strategy proposed in this work to reinforcement learning based methods.

%% file: appendix.tex
\section*{Appendix}

\begin{figure}[t]
    \begin{center}
    \includegraphics[width=\linewidth]{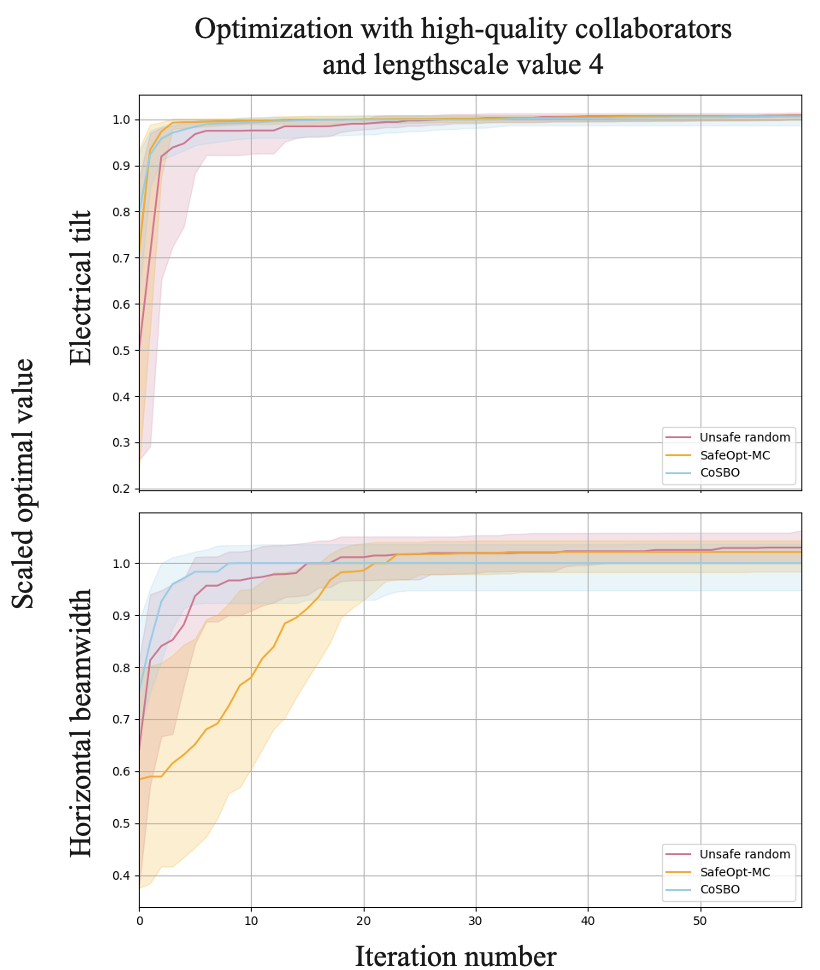}
    \caption{Optimization method comparison for tilt and beamwidth using high-quality collaborators and a GP model with lengthscale value \qty{4}. Curves correspond to the optimum obtained at each iteration of the process for a budget of 60 iterations. The shaded areas represent the \qty{25}{\percent}-\qty{75}{\percent} inter-quartile range of values obtained over the \qty{15} simulated networks times the \qty{10} different starting points. CoSBO method is displayed in blue, SafeOpt-MC in orange, and an unsafe random exploration method in violet red.}
    \label{fig:optims-l4-pi}        
    \end{center}
\end{figure}

\begin{figure}[t]
    \begin{center}
    \includegraphics[width=.5\textwidth]{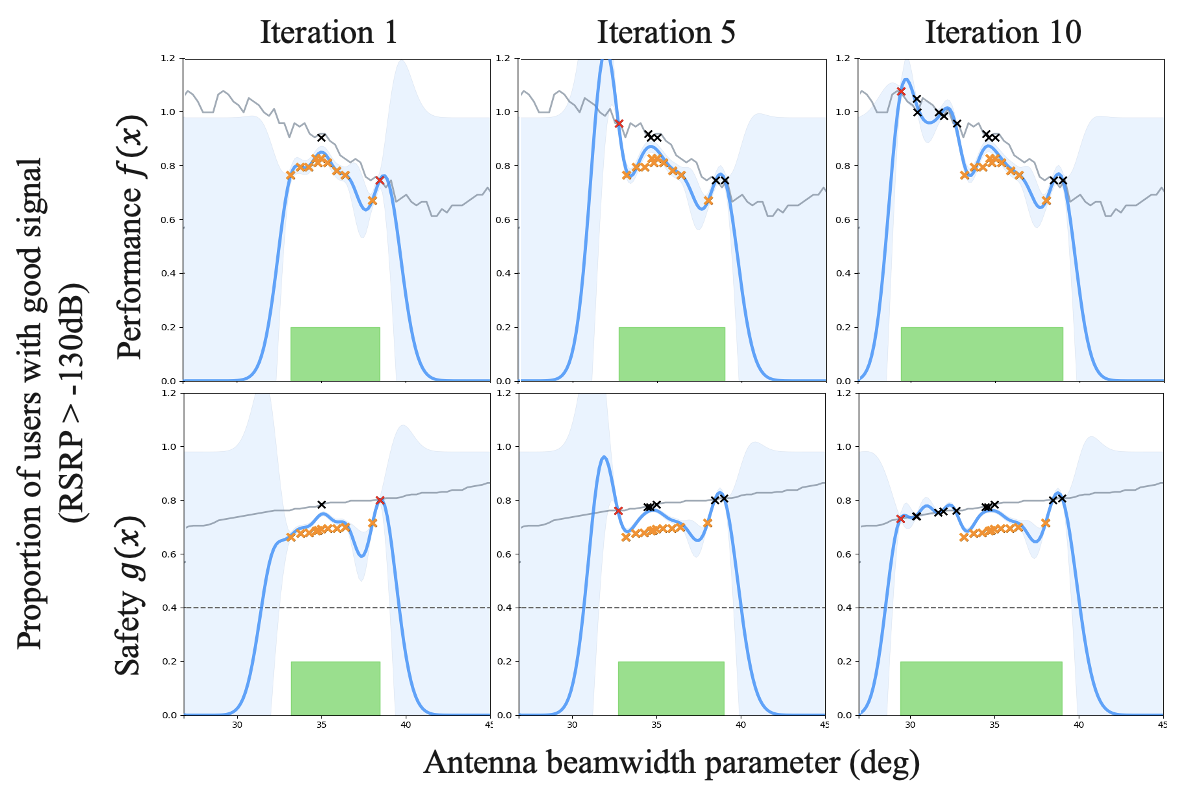}
    \caption{Optimization process of the horizontal beamwidth of an antenna using the CoSBO algorithm to maximize the performance function (gray curve on the top) while satisfying the safety constraint (gray curve on the bottom). The visualization shows the progress of the estimated functions (light blue curves) at the first, fifth, and tenth iterations using a smooth GP model with lengthscale \qty{4}. The algorithm starts with collaborator data (orange crosses) that is added as observed data (black crosses). Then, based on the GP posterior and its confidence intervals (light blue areas), it iteratively selects and evaluates new data (red crosses) that are above the safety threshold (gray dashed line) and within the current safe set (green set).}
    \label{fig:bw_iterations_l4}        
    \end{center}
\end{figure}

To further understand the behavior of both safe algorithms, \textsc{CoSBO} and \textsc{SafeOpt-MC} \cite{berkenkamp2023bayesian}, we conducted additional experiments by modifying a key GP hyperparameter: the kernel lengthscale \begin{math}\ell\end{math}, which influences the smoothness of the estimated functions \begin{math}\hat f, \hat g\end{math}. We observe the effect of a larger lengthscale value (\begin{math}\ell = 4\end{math}) compared to the standard configuration used previously (\begin{math}\ell = 1\end{math}). The smoothness of the GP posterior increases with the lengthscale, which also affects the width of the confidence intervals used to guide the exploration \cite{rasmussenwilliams2006gaussian}. Smoother models produce narrower intervals near the observations, encouraging the selection of points farther from the known data. This leads to increased exploration, but also raises the risk of selecting unsafe points due to the wider safe region. \Cref{fig:bw_iterations_l4} presents additional results for the optimization of the horizontal beamwidth parameter, similar to those in \cref{fig:bw_iterations} of the main text, but using the same topological map under a different traffic volume. The plot illustrates an increase in smoothness in the estimated function. Analogously to \cref{fig:optdefaultweak}, in \cref{fig:optims-l4-pi} we measure the sample efficiency of the methods using a GP model with lengthscale 4. The results indicate that smoother GPs better capture the tilt parameter functions (top), which are inherently smooth, while slightly decreasing the performance on the beamwidth functions (bottom), which are highly variable. For the beamwidth optimization problem, we can see that our method still provides significant improvements in sample efficiency.